# Recognizing three-dimensional phase images with deep learning

Weiru Fan[1,*], Tianrun Chen[1,*], Xingqi Xu[1], Ziyang Chen[2,†], Huizhu Hu[1,3], Delong Zhang[1,†], Da-Wei Wang[1,†], Jixiong Pu[2] and Shi-Yao Zhu[1]

[1]Interdisciplinary Center for Quantum Information and State Key Laboratory of Modern Optical Instrumentation, Zhejiang Province Key Laboratory of Quantum Technology and Device, and Department of Physics, Zhejiang University, Hangzhou, Zhejiang 310027, China
[2]College of Information Science and Engineering, Fujian Key Laboratory of Light Propagation and Transformation, Huaqiao University, Xiamen, Fujian 361021, China
[3]Zhejiang Laboratory, Hangzhou, Zhejiang 311121, China

*These authors contributed equally: Weiru Fan, Tianrun Chen
†Email: ziyang@hqu.edu.cn; dlzhang@zju.edu.cn; dwwang@zju.edu.cn

**Optical phase contains key information for biomedical and astronomical imaging. However, it is often obscured by layers of heterogeneous and scattering media, which render optical phase imaging at different depths an utmost challenge. Limited by the memory effect, current methods for phase imaging in strong scattering media are inapplicable to retrieving phases at different depths. To address this challenge, we developed a speckle three-dimensional reconstruction network (STRN) to recognize phase objects behind scattering media, which circumvents the limitations of memory effect. From the single-shot, reference-free and scanning-free speckle pattern input, STRN distinguishes depth-resolving quantitative phase information with high fidelity. Our results promise broad applications in biomedical tomography and endoscopy.**

Complementing conventional brightfield imaging, phase imaging is not limited to absorption-based contrast and provide abundant information for applications in biomedical endoscopy. However, random media such as ground glasses, biological tissues, and multimode fibers (MMF), are significant obstacles in biomedical imaging, since they scramble spatial information in random diffusion[1-4]. To solve this problem, many methods have been developed to retrieve the information encoded in the random speckle by using ballistic photons[5,6], wavefront shaping[7,8], transmission matrix[9-11] and machine learning[12-16]. Instead of eliminating the obstacles, other methods take advantages of the distortion by using random media as quasi-optical lenses[17,18]. An example is the lens-free imaging based on speckle intensity correlation techniques, which rely on the (angular) memory effect[19], i.e., the correlation and shift-invariance of speckle patterns within a certain range of illumination angle. However, the infinite focal distance of this lateral (angular) memory-effect-based speckle correlation technique restricts the system in two-dimensional (2D) imaging only.

Recently, the intensity-correlation-based technique has also been applied to three-dimensional (3D) imaging in presence with strong scattering medium[20-23]. The early practice of 3D imaging is to use multi-view stereo technique as a non-invasive method, which captures multiple 2D projections from multiple views and reconstructs a 3D model[24-26]. Complicated optical setup and/or time-consuming scanning process were developed to extend memory effect from 2D to 3D, achieving 3D imaging through strongly scattering media[20-22]. These methods have difficulties in addressing the challenges of depth resolution and large field-of-view imaging, because axial shift invariance could only retrieve the shape of an object in a limited range of illumination angle. In addition, an incoherent illumination is required for single-shot measurement, so that only the amplitude contrast can be obtained[17,20-22]. The critical information of examining transparent sample, the optical phase, is difficult to retrieve. Other methods either require complex experimental setup with a reference beam, or special statistical properties of the scattering media or less-scalable computational processes for objects at multiple distances[27-31]. One way to avoid the problems is to use computational imaging techniques such as deep learning[12-16,31-33]. However, these methods have only been applied in 2D imaging or without random media. Retrieving various information from multiple depths, especially with spatially overlapping regions between different layers, remains challenging.

Here, we developed a single-shot multilayer phase imaging approach to achieve 3D phase imaging in random media (Fig. 1), termed as speckle three-dimensional reconstruction network (STRN). The challenge that we address here is the 3D information retrieval from a projected 2D image without any preprocessing or prior knowledge. Deep learning-based methods have shown its capability in finding out the statistical characteristics of modeling a specific process. Here the key is to model the inverse physical process in order to reconstruct the 3D phase information from the 2D image speckle pattern. However, solving this inverse problem is hampered by under-sampling, since the explicit information contained in the 2D speckle image is always less than that in the three 2D phase images. We show that such a gap can be filled by neural network training with large number of matched image pairs. The STRN circumvent the limitations of memory-effect-based methods through a delicate neural network. With a raw captured speckle image as the input, STRN reconstructs three layers of phase images at different depths with high fidelity, which is promising for applications in depth-resolving biomedical phase imaging.

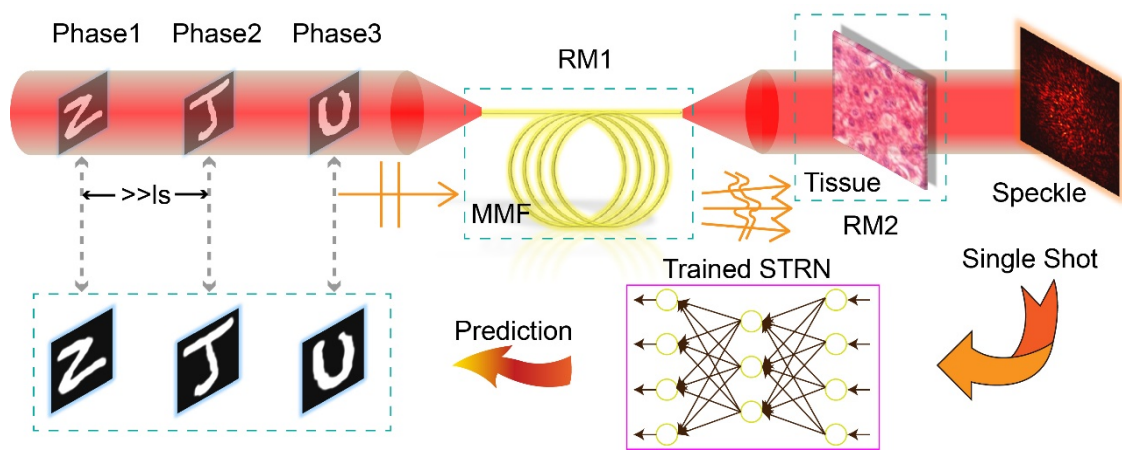

Fig. 1. 3D phase reconstruction from 2D speckle pattern. Light carrying 3D phase information goes through different random media and generates a complex speckle. Spatial light modulators are used to impose phase masks on the light. The free-space propagation length between two SLMs is much larger than the memory effect range (ls). The acquired speckle is input into a trained neural network to retrieve original 3D phase images.

To mimic a real-world scenario without the memory effect, a multimode fiber (MMF) and a biological tissue were sequentially combined to generate the speckle pattern. The MMF only has rotational memory effect while the biological tissue merely has angular memory effect[34,35]. When used in concatenate, they eliminate the 2D memory effect. We also arrange the distances between each layer of phase images large enough to eliminate the 3D memory effect. As a result, the generated 2D speckle images are free from the memory effect, mimicking the challenge in retrieving multilayer phase images in biomedical imaging. To solve this problem, we trained STRN by feeding ground truth with corresponding phase image sets (See methods) and tested it with unseen image sets. Pearson correlation coefficient (PCC) was used to quantitatively evaluate the output by comparing reconstructed phase images with ground truth images. We recovered phase images at different depths by feeding a single speckle image into the neural network, and thus we realized single shot 3D phase imaging. It took STRN only 1 second to process each speckle image to generate the three phase images on consumer-grade graphics cards.

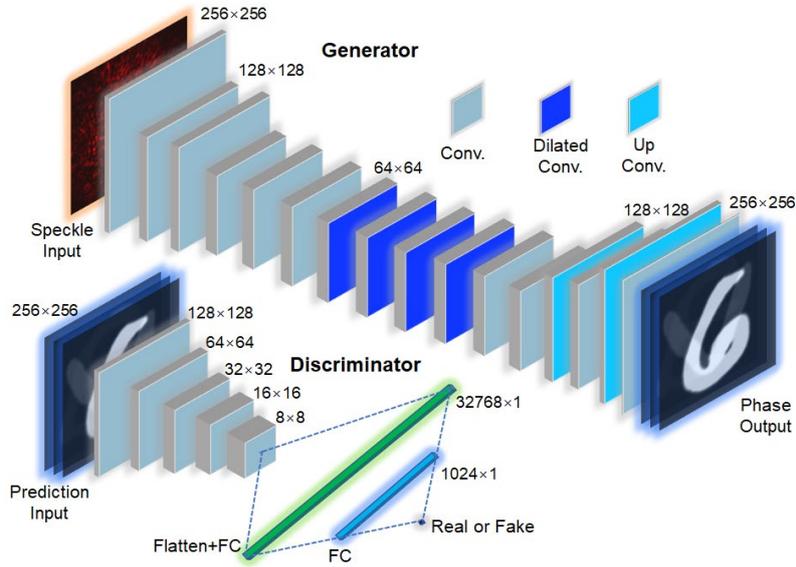

Fig. 2. Schematic of STRN. The network consists of a generator network and an auxiliary discriminator network.

The design and development of the STRN was inspired by generative adversarial network (GAN)[36], a powerful tool to accomplish the inter-domain image transformation from one to multiple domains. GAN has been applied to imaging systems such as data enhancement and multi-modal image transformation[37,38]. Based on previous studies on generative adversarial network, STRN consists of a "generator" and a "discriminator" (Fig. 2). The generator is an encode-decode structured neural network that generates reconstructed phase images from the input speckle images. The generated images go through the discriminator that distinguishes the "real" or "fake" images. The outputs of the discriminator are used to adjust the loss function, forcing the generator to generate desired outputs. The generator has downsampling and upsampling processes. The downsampling process reduces the resolution of the image, which saves the computation and memory cost and forces the network to capture the prominent features of samples. To avoid the information loss, strided convolutions instead of widely-used pooling operations are employed, which is critical in obtaining more information for further processes. Due to the obscured structural relation between the input and output images, more spatial information is needed to acquire high-quality image reconstruction. We insert multiple dilated convolution layers to "see" larger regions in the images[39]. The training process is a min-max optimization problem. The generator and discriminator networks are jointly adjusted in each iteration until the discriminator cannot distinguish the generated images, meaning that the 3D phase information is successfully recovered.

The performance of the STRN reconstruction of the 3D phase information is shown with predicted phase patterns in Fig.3 (a) and quantitative evaluation in Fig.3 (b). The reconstruction qualities of the three phase images are slightly different, which is probably due to the propagation and the diffraction of the light. The dark-spot regions in the intensity distribution are caused by shadowing between different phase image layers, resulting in the loss of the modulation efficiency in hidden layers, such that the phase masks are only partially loaded. Moreover, to mimic the 3D memory effect, the planes are separated by meters, so that the diffraction-induced incomplete modulation is magnified. It is expected that with shorter distance between layers in real applications the variation of reconstruction qualities of different layers will be reduced.

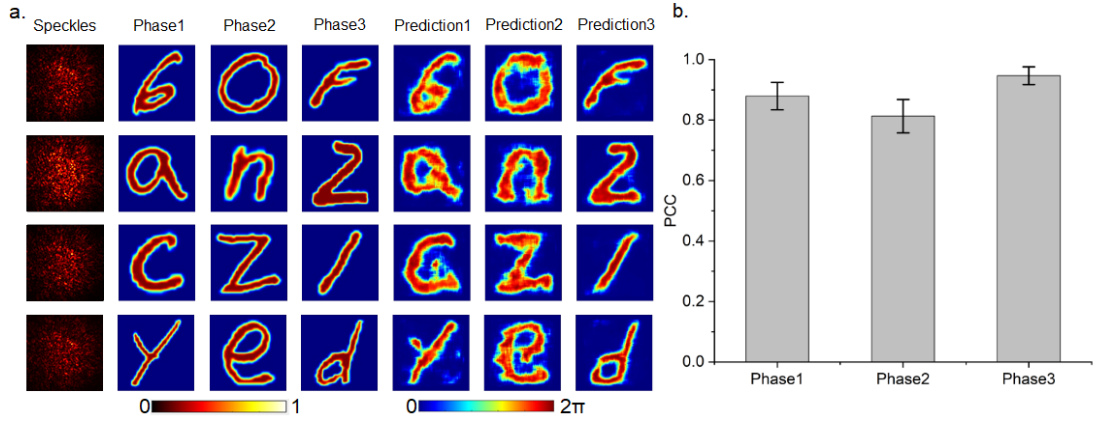

Fig. 3. Quantitative evaluation of the performance of STRN for reconstruction of 3D phase information. (a) Results of reconstruction of 3D phase information from 2D speckle; (b) Performance evaluation of the three reconstructed phase images. Error Bar: ± standard deviation. PCC: Pearson correlation coefficient.

The implementation is a preliminary attempt in recognizing multiple stacked phase objects by single-shot measurement. The problem of image overlapping is solved by separating individual phase shifts into different channels on the decoder path. In real-world, certain morphological correlations can appear between multiple phase layers. The inherent phase correlation allows information sharing in neural network channels[40], which can process information more efficiently than using part of them to retrieve an individual output. Although we only demonstrated the reconstruction of three phase images, the approach can be developed with moderate modification to cope with more phase images.

Imaging techniques based on the memory effect are mostly effective in thin scattering layers. The STRN is applicable to thick and multiple layered scattering media. Our approach achieves lensless multi-depth 3D phase imaging by using the diffusive layer as a lens-like imaging element, which can be used to image complex targets with a large field of view. Since it does not rely on any memory effects, the phase image reconstruction can be achieved from a single shot after the training is completed. This allows instantaneous imaging of dynamic objects through random media.

Moreover, imaging multiple layers with a single shot image addresses a common problem in various areas such as autonomous driving, chip detection, and biological tissue chromatography[29,31,41]. Our work provides a general solution to such problems, notably on how to acquire training data sets for a dedicated learning network. To date, the most effective way to solve this problem is the transfer learning[42-44]. It allows us to take training data from a simpler system and combine it with little or no data from the current system for training. This capability allows us to extend STRN to a dynamic diffuser or a turbulent medium.

In conclusion, we demonstrated a novel imaging scheme for retrieving 3D phase information from 2D speckle patterns. A deep learning network, STRN, is designed to enable the 3D phase imaging scheme. Recognizing overlapping phase images in random media without any reference or scanning process, this novel scheme has potential applications in astronomical and biological imaging.

**Method**
1. Experimental setup

The experimental setup is shown in Fig. S1. The laser beam from stabilized He-Ne laser (HNL150L-EC, Thorlabs) goes through a microscopic objective (20×, NA=0.4), a pinhole (20μm) and a lens (f=50 mm). The light spot on the three SLMs (Pluto-VIS, Holoeye, pixel size= 8μm, 1080×1920 pixels) has a diameter of 3 mm. The phases are modulated with a cross section of 256×256 pixels, simulating three cross sections of a 3D phase object. The three phase images are placed one after another with a separation of about 80 cm, assuring that the memory effect is absent. The first and the third phase images are generated from different areas in SLM1, while the second phase image is generated by SLM2. The light passes through complex random media that consist of an MMF and a biological sample (a two-layer tissue slice of mouse liver with 20 μm thickness for each layer), generating complex speckle patterns. The generated speckle is recorded by a charge coupled device (CCD, GT1910, Prosilica, AVT), with a spatial resolution of 512×512 pixels. The dataset includes 120,000 speckle patterns and 360,000 phase masks, with 90% of them being used as the training set and the rest 10% for testing.

2. Implementation of STRN

**Structure**: STRN consists of a generator and a discriminator. The generator produces the predicted images by an encoder and a decoder. The input single channel 256 × 256 images go through a composite layer with 5 × 5 convolution and result in a 64-channel feature map. Then, the feature maps are compressed to a lower resolution and the channel number is doubled. Each compression is performed via a composite layer with 3 × 3 convolution kernels with a stride 2, followed by a dropout layer with a dropout rate of 0.8, and another composite layer with 3 × 3 convolution kernels with a stride 1. Each composite layer includes a convolution operation, a batch normalization, and a Leaky ReLu activation function.

The feature maps then go through a decoding process, where layers are operated by the opposite operations of the encoding process, i.e., a composite layer with 3 × 3 convolution with a stride 1, a composite layer with 4 × 4 deconvolution with a stride 1/2, followed by a dropout layer with a dropout rate of 0.8. Each composite layer includes a convolution operation, a batch normalization, and a Leaky ReLu activation function. The final output stage is a composite layer with the input of 32-channel 256×256 feature maps going through 3 × 3 convolution with a stride 1 and a tanh activation function, generating 3-channel 256×256 images as the output. The output image goes through a discriminator, in which a series of downsampling operations are performed by multiple 5 × 5 convolutions with a stride 2, resulting in 512-channel 8×8 feature maps. The pixels of images are rearranged to a flat array, and go through a fully connected layer, generating a single true-or-false output.

**Training**: The implementation of the network uses the Tensorflow framework. After the network is built, the training and evaluation processes are performed on a server (NVIDIA Tesla V100, CUDA 10.2). Handwritten characters from the Extended Modified National Institute of Standards and Technology (EMNIST) dataset and Quick Draw dataset are used as the SLM input and the ground truth image for STRN. For every collected dataset, 90% and 10% are used for training and testing, respectively. The training set is fed into the STRN with 100 epochs for the generator first. With the generator being pre-trained, the discriminator and generator are jointly trained for 310 epochs. The learning rate of Adam optimizer is set to be 0.0008. When the STRN is sufficiently trained, the test set is fed to STRN to evaluate its performance by Pearson Correlation Coefficient (PCC). More specific details can be found in the Supplementary.

**Data availability**. All necessary data are shown in the figures of the manuscript. Further data will be provided upon reasonable request.

## References
1. J. Bertolotti, E. G. van Putten, C. Blum, A. Lagendijk, W. L. Vos, and A. P. Mosk, Non-invasive imaging through opaque scattering layers, Nature, 491, 232-234 (2012).
2. P. Sebbah, Waves and imaging through complex media, (Kluwer, 2001).
3. I. M. Vellekoop, and A. P. Mosk, Focusing coherent light through opaque strongly scattering media, Opt. Lett., 32(16), 2309-2311 (2007).
4. B. Redding, S. F. Liew, R. Sarma, and H. Cao, Compact spectrometer based on a disordered photonic chip, Nat. Photon., 7, 746-751 (2013).
5. D. Huang, E. A. Swanson, C. P. Lin, J. S. Schuman, W. G. Stinson, W. Chang, M. R. Hee, T. Flotte, K. Gregory, C. A. Puliafito, and J. G. Fujimoto, Optical coherence tomography, Science, 254, 1178-1181 (1991).
6. A. Velten, T. Willwacher, O. Gupta, A. Veeraraghavan, M. G. Bawendi, and R. Raskar, Recovering three-dimensional shape around a corner using ultrafast time-of-flight imaging, Nat. Commun., 3, 745 (2012).
7. O. Katz, E. Small, and Y. Silberberg, Looking around corners and through thin turbid layers in real time with scattered incoherent light, Nat. Photon., 6, 549-553 (2012).
8. J. Park, J. Park, K. Lee, Y. Park, Disordered optics: exploiting multiple light scattering and wavefront shaping for nonconventional optical elements, Adv. Mater., 32(35), 1903457 (2019).
9. S. Popoff, G. Lerosey, M. Fink, A. C. Boccara, and S. Gigan, Image transmission through an opaque material, Nat. Commun., 1, 81 (2010).
10. M. Plöschner, T. Tyc, and T. Čižmár, Seeing through chaos in multimode fibres, Nat. Photon., 9, 529-535 (2015).
11. Y. Choi, T. D. Yang, C. Fang-Yen, P. Kang, K. J. Lee, R. R. Dasari, M. S. Feld, and W. Choi,

Overcoming the diffraction limit using multiple light scattering in a highly disordered medium, Phys. Rev. Lett., 107, 023902 (2012).
12. Y. Li, Y. Xue, and L. Tian, Deep speckle correlation: a deep learning approach toward scalable imaging through scattering media, Optica, 5(10), 1181-1190 (2018).
13. S. Li, M. Deng, J. Lee, A. Sinha, and G. Barbastathis, Imaging through glass diffusers using densely connected convolutional networks, Optica, 5(7), 803-813 (2018).
14. B. Rahmani, D. Loterie, G. Konstantinou, D. Psaltis, C. Moser, Multimode optical fiber transmission with a deep learning network, Light Sci. Appl., 7(1), 69 (2018).
15. N. Borhani, E. Kakkava, C. Moser, D. Psaltis, Learning to see through multimode fibers, Optica, 5(8), 960–966 (2018).
16. G. Barbastathis, A. Ozcan, and G. Situ, On the use of deep learning for computational imaging, Optica, 6, 921-943 (2019).
17. O. Katz, P. Heidmann, M. Fink, and S. Gigan, Non-invasive single-shot imaging through scattering layers and around corners via speckle correlations, Nat. Photon., 8, 784-790 (2014).
18. I. Freund, Looking through walls and around corners, Physica A, 168(1), 49-65 (1990).
19. I. Freund, M. Rosenbluh, and S. Feng, Memory effects in propagation of optical waves through disordered media, Phys. Rev. Lett., 61, 2328 (1988).
20. A. K. Singh, D. N. Naik, G. Pedrini, M. Takeda, and W. Osten, Exploiting scattering media for exploring 3D objects, Light Sci. Appl., 6(1), e16219 (2016).
21. Y. Okamoto, R. Horisaki, and J. Tanida, Noninvasive three-dimensional imaging through scattering media by three-dimensional speckle correlation, Opt. Lett., 44(10), 2526-2529 (2019).
22. W. Li, J. Liu, S. He, L. Liu, and X. Shao, Multitarget imaging through scattering media beyond the 3D optical memory effect, Opt. Lett., 45(10), 2692-2695 (2020).
23. M. Takeda, A. K. Singh, D. N. Naik, G. Pedrini, and W. Osten, Holographic correloscopy-unconventional holographic techniques for imaging a three-dimensional object through an opaque diffuser or via a scattering wall: a review, IEEE Trans. Industr. Inform., 12(4), 1631-1640 (2016).
24. P. J. Shaw, D. A. Agard, Y. Hiraoka, and J. W. Sedat, Tilted view reconstruction in optical microscopy. Three-dimensional reconstruction of Drosophila melanogaster embryo nuclei, Biophys. J., 55(1), 101-110 (1989).
25. J. Swoger, P. Verveer, K. Greger, J. Huisken, and E. H. K. Stelzer, Multi-view image fusion improves resolution in three-dimensional microscopy, Opt. Express, 15(13) 8029-8042 (2007).
26. J. Sharpe, U. Ahlgren, P. Perry, B. Hill, A. Ross, J. Hecksher-Sørensen, R. Baldock, and D. Davidson, Optical Projection Tomography as a Tool for 3D Microscopy and Gene Expression Studies, Science, 296(5567), 541-545 (2002).
27. N. Antipa, G. Kuo, R. Heckel, B. Mildenhali, E. Bostan, R. Ng, and L. Waller, DiffuserCam: lensless single-exposure 3D imaging, Optica, 5(1), 1-9 (2018).
28. S. Kodama, M. Ohta, K. Ikeda, Y. Kano, Y. Miyamoto, W. Osten, M. Takeda, and E. Watanabe, Three-dimensional microscopic imaging through scattering media based on in-line phase-shift digital holography, Appl. Opt., 58(34), G345-G350 (2019).
29. T. Kim, R. Zhou, M. Mir, S. D. Babacan, P. S. Carney, L. L. Goddard, and G. Popescu, White-light diffraction tomography of unlabeled live cells, Nat. Photon., 8, 256-263 (2014).
30. T. H. Nguyen, M. E. Kandel, M. Rubessa, M. B. Wheeler, and G. Popescu, Gradient light interference microscopy for 3D imaging of unlabeled specimens, Nat. Commun., 8, 210 (2017).

31. A. Goy, G. Rughoobur, S. Li, K. Arthur, A. I. Akinwande, and G. Barbastathis, High-resolution limited-angle phase tomography of dense layered objects using deep neural networks, Proc. Natl. Acad. Sci., 116(40), 19848-19856 (2019).
32. G. Carleo, I. Cirac, K. Cranmer, L. Daudet, M. Schuld, N. Tishby, L. Vogt-Maranto, and L. Zdeborová, Machine learning and the physical sciences, Rev. Mod. Phys., 91(4), 045002 (2019).
33. E. Nehme, D. Freedman, R. Gordon, B. Ferdman, L. E. Weiss, O. Alalouf, T. Naor, R. Orange, T. Michaeli, and Y. Shechtman, DeepSTORM3D: dense 3D localization microscopy and PSF design by deep learning, Nat. Met., 17(7), 734-740 (2020).
34. S. Schott, J. Bertolotti, J. Léger, L. Bourdieu, S. Gigan, Characterization of the angular memory effect of scattered light in biological tissues, Opt. Express, 23, 13505-13516 (2015).
35. L. V. Amitonova, A. P. Mosk, P. W. H. Pinkse, Rotational memory effect of a multimode fiber, Opt. Express, 23, 20569-20575 (2015).
36. I. J. Goodfellow, J. Pouget-Abadie, M. Mirza, B. Xu, D. Warde-Farley, S. Ozair, A. Courville, Y. Bengio, Generative adversarial nets, Adv. Neural Inf. Process. Syst., 27, 2672-2680 (2014).
37. D. Li, Lin Shao, B. Chen, X. Zhang, M. Zhang, B. Moses, D. E. Milkie, J. R. Beach, J. A. Hammer, M. Pasham, T. Kirchhausen, M. A. Baird, M. W. Davidson, P. Xu, and E. Betzig, Extended-resolution structured illumination imaging of endocytic and cytoskeletal dynamics, Science, 349(6251), aab3500 (2015).
38. G. Litjens, T. Kooi, B. E. Bejnordi, A. A. A. Setio, F. Ciompi, M. Ghafoorian, J. A. W. M. van der Laak, B. van Ginneken, C. I. Sánchez, A survey on deep learning in medical image analysis, Med. Image Anal., 42, 60-88 (2017).
39. F. Yu, and K. Vladlen, Multi-scale context aggregation by dilated convolutions, arXiv, 1511.07122 (2015).
40. X. Han, H. Laga, M. Bennamoun, Image-based 3D object reconstruction: state-of-the-art and trends in the deep learning era, arXiv, 1906.06543 (2019).
41. X. Chen, H. Ma, J. Wan, B. Li, T. Xia, Multi-view 3D object detection network for autonomous driving, arXiv: 1611.07759 (2017).
42. S. Goswami, C. Anitescu, S. Chakraborty, T. Rabczuk, Transfer learning enhanced physics informed network for phase-field modeling of fracture, Theor. Appl. Fract. Mech., 106, 102447 (2020).
43. V. Cheplygina, M. de Bruijne, J. P. W. Pluim, Not-so-supervised: A survey of semi-supervised, multi-instance, and transfer learning in medical image analysis, Med. Image Anal., 54, 280-296 (2019).
44. H. Shin, H. R. Roth, M. Gao, L. Lu, Z. Xu, I. Nogues, J. Yao, D. Mollura, R. M. Summers, Deep convolutional neural networks for computer-aided detection: CNN architectures, dataset characteristics and transfer learning, IEEE T. Med. Imaging, 35(5), 1285-298 (2016).

# Supporting Information

## Transmission Model

According to angular spectrum-based theorem, the phase loading and the transmission of light can be calculated as

$$E_{l+1} = \mathcal{F}^{-1}\left\{\mathcal{F}(E_l \cdot P_l) e^{-i\frac{2\pi}{\lambda}\left(\sqrt{1-(\lambda\xi)^2-(\lambda\eta)^2}\right)\Delta z}\right\} \quad (S1)$$

where $P_l$ is the phase map imposed by the SLM, $E_l$ is the electric field in front of the SLM, $\lambda$ is the wavelength, $\eta$ and $\xi$ are the two-dimensional spatial coordinates, $\Delta z$ is the distance between two SLMs, $\mathcal{F}$ and $\mathcal{F}^{-1}$ denote the Fourier transform and inverse Fourier transform, and $l$ is the number of the layers.

The scattering process can be expressed as

$$E_m^{s+1} = \sum_{n=1}^{N} t_{mn}^s E_n^s, \quad (S2)$$

where $t_{mn}^s$ represents the elements of trnasmission matrix of the $s$'th scattering media, $E_n^s$ is the electric field of the $n$'th input element, and $E_m^{s+1}$ is the electric field of the $m$'th output field.

## Mathematical Model of STRN

STRN uses paired samples (2D speckle and 3D phase information) of the experimental data for training. After well-trained, the transformation from 2D speckles to 3D phases can be realized. For simplicity, we denote the light propagation and transformation with a forward operator $T$, and thus derived the computation process as follows[13,31]

$$\hat{w} = \arg\min_{w}\left\{\|r - Tw\|^2 + \delta(w)\right\} \quad (S3)$$

where, $w$ denotes the unknown 3D object, $r$ is the generated speckle, $\delta$ is a regularization term to eliminate the influence of the experimental noise. The optimization process of STRN is to minimize the value of the function by continuously adjusting the parameters of the neural network, derived as

$$q = \arg\min \sum_{n=1}^{N}\left\|g_n - \prod_{\theta} H_\theta \times f_n\right\|^2 + \Delta \quad (S4)$$

where $q$ represents the objective function for evaluating the network performance and suggesting the adjustment of parameters, $f_n$ and $g_n$ are the input and output of STRN, $\Delta$ is a regularization term to avoid the overfitting, and $H_\theta$ denotes the $\theta$'th mathematical operation or transformation of the network. STRN minimizes the objective function by adjusting the parameters of $H_\theta$ with abundant data. Finally, STRN acquires the ability to predict unknown data.

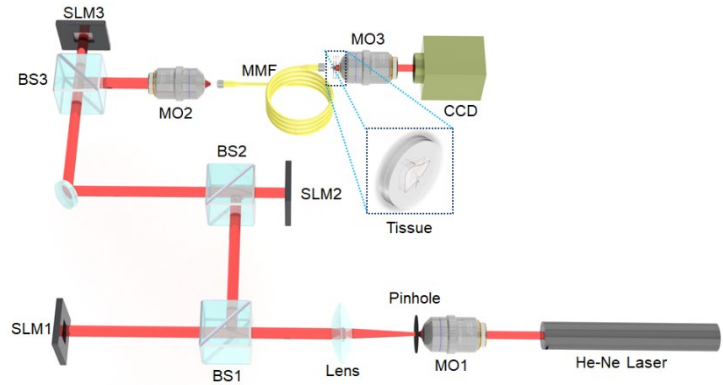

Fig.S1. Experimental setup. MO: microscopy objective; BS: beam splitter; SLM: spatial light modulator; M: mirror; MMF: Multimode fiber; SF: spatial filter.

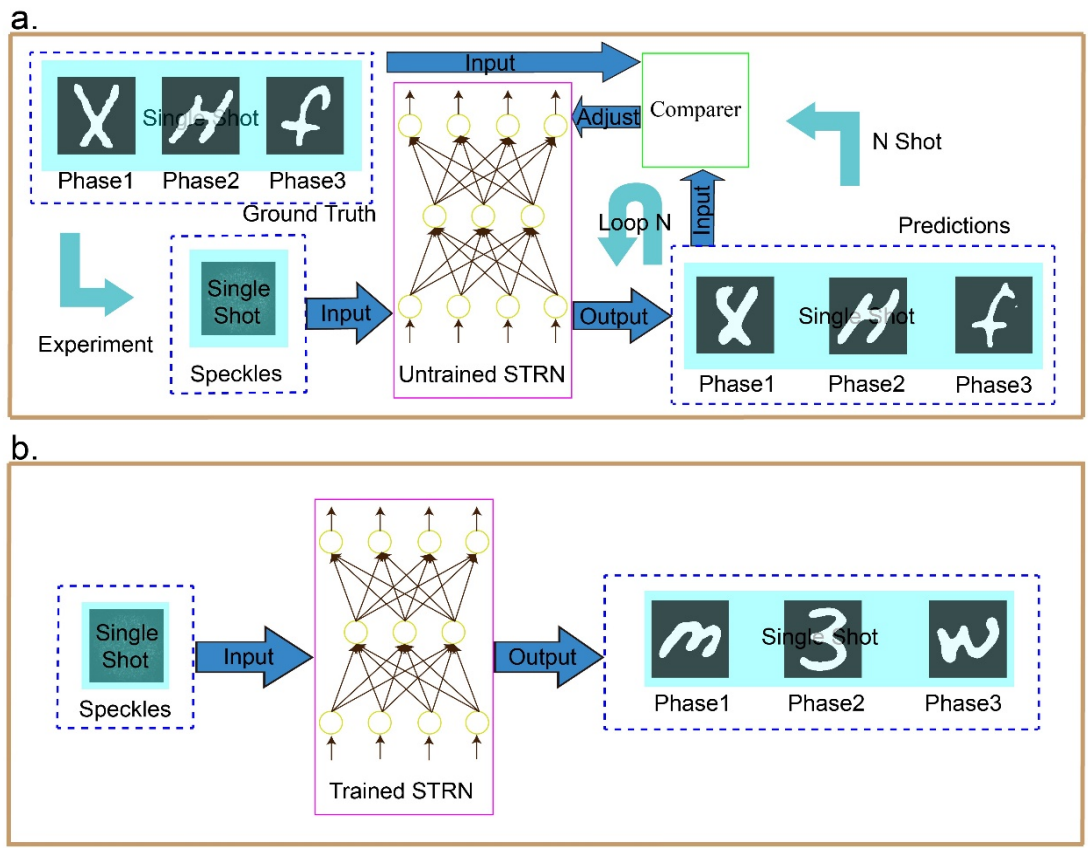

Fig. S2. Overview of STRN. (a) Network training by the collected data; (b) Phase reconstruction from the detected speckle by trained network.

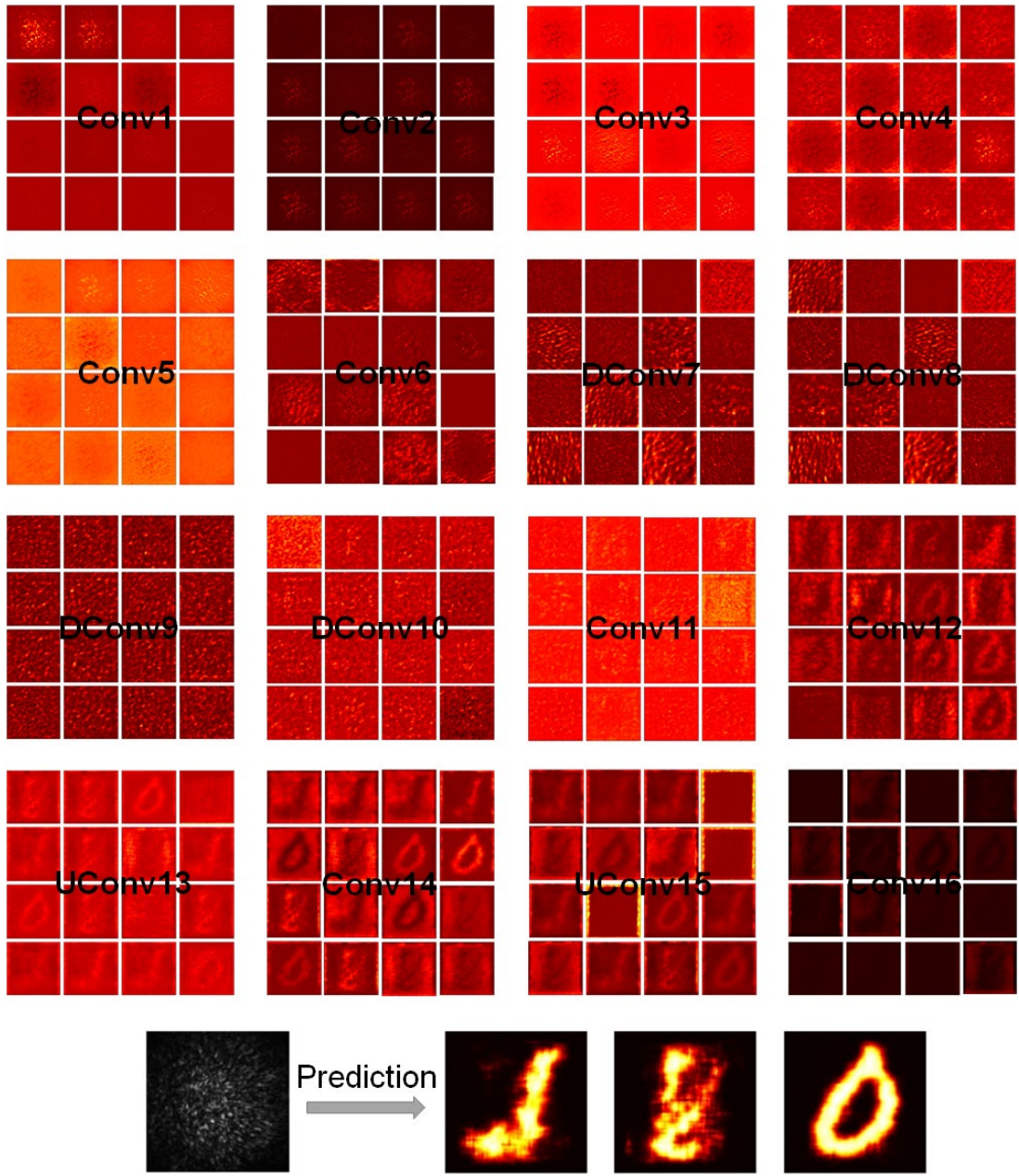

Fig. S3. The intermediate activation maps in generator. Conv: convolution; DConv: Dilated convolution; UConv: Up Convolution.

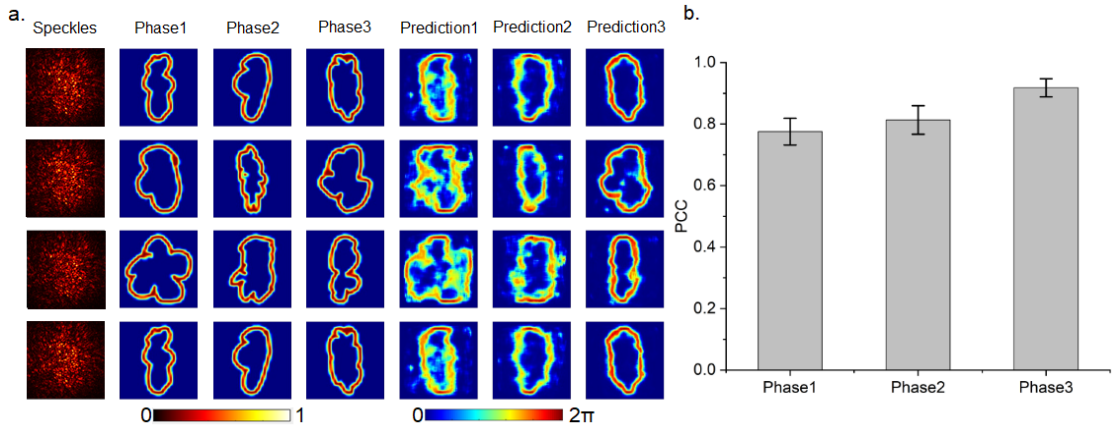

Fig. S4. The result of cloud in Quick, Draw! dataset.

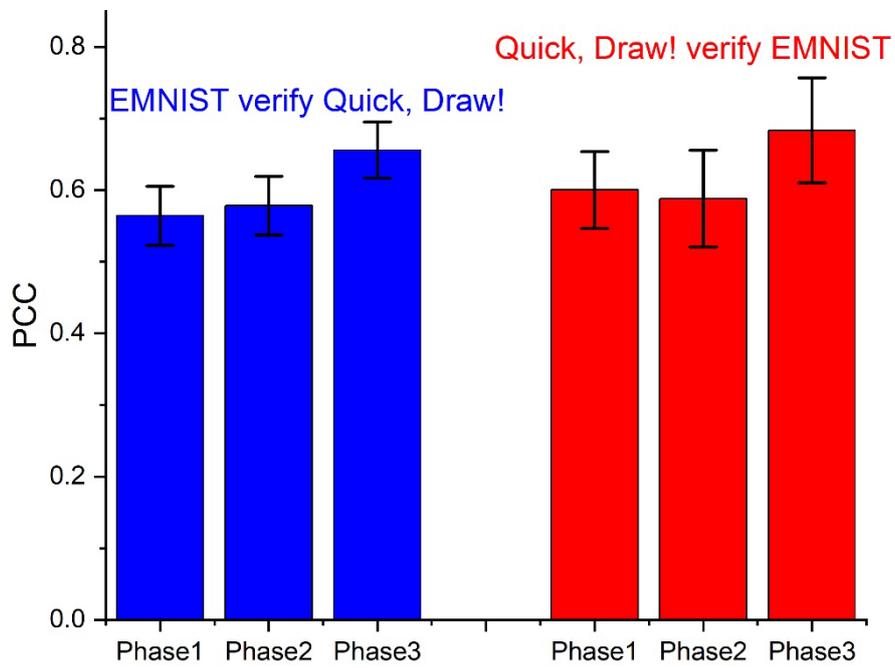

Fig. S5. Cross-verify

Tab.1. Parameters of generator for STRN.

| Layer | Composition | Num of Parameter | |
|---|---|---|---|
| Conv1 | 5x5x1x64 | 1600 | 128 |
| Conv2 | 3x3x64x128 | 73728 | 256 |
| Conv3 | 3x3x128x128 | 147456 | 256 |
| Conv4 | 3x3x128x256 | 294912 | 512 |
| Conv5 | 3x3x256x256 | 589824 | 512 |
| Conv6 | 3x3x256x256 | 589824 | 512 |
| Dilated conv1 | 3x3x256x256 | 589824 | 512 |
| Dilated conv2 | 3x3x256x256 | 589824 | 512 |
| Dilated conv3 | 3x3x256x256 | 589824 | 512 |
| Dilated conv4 | 3x3x256x256 | 589824 | 512 |
| Conv7 | 3x3x256x256 | 589824 | 512 |
| Conv8 | 3x3x256x257 | 589824 | 512 |
| Upconv1 | 4x4x128x256 | 524288 | 256 |
| Conv9 | 3x3x128x128 | 147456 | 256 |
| Upconv2 | 4x4x64x128 | 131072 | 128 |
| Conv10 | 3x3x64x32 | 18432 | 64 |
| Conv11 | 3x3x32x3 | 864 | 0 |
| Total size: 6064352 | | | |

Tab.2. Parameters of discriminator for STRN.

| Layer | Composition | Num of Parameter | |
|---|---|---|---|
| Conv1 | 5x5x3x64 | 4800 | 128 |
| Conv2 | 5x5x64x128 | 204800 | 256 |
| Conv3 | 5x5x128x256 | 819200 | 512 |
| Conv4 | 5x5x256x512 | 3276800 | 1024 |
| Conv5 | 5x5x512x512 | 6553600 | 1024 |
| FC1 | 32768x1024 | 33554432 | 1024 |
| FC2 | 1024x1 | 1024 | 1 |
| Total size: 44418625 | | | |